\documentclass[12pt]{article}
\usepackage{graphicx}
\usepackage{epsfig}
\usepackage{amsmath}
\usepackage{color}

\makeatletter
\@addtoreset{equation}{section}

\renewcommand{\thefootnote}{\fnsymbol{footnote}}
\makeatother
\newcommand {\beq}{\begin{eqnarray}}
\newcommand {\eeq}{\end{eqnarray}}
\newcommand {\non}{\nonumber\\}

\newcommand{\spone}{0.9}  

\newcommand{\sptwo}{1.4}
\newcommand{\spthree}{2.4}

\newcommand{\singlespace}{\edef\baselinestretch{\spone}\Large\normalsize}

\newcommand{\doublespace}{\edef\baselinestretch{\sptwo}\Large\normalsize}
\newcommand{\threespace}{\edef\baselinestretch{\spthree}\Large\normalsize}

\singlespace
\threespace
\doublespace

\everymath={\displaystyle}
\begin{document}
\doublespace

~
\vspace{2cm}

\begin{center}
{\bf {\Large Non-Abelian Vortex-String Dynamics from Nonlinear Realization
}\\
$~$\\
Lu-Xin Liu$^a$  {\hspace{3pt}} and {\hspace{3pt}}  Muneto Nitta$^b$ 
\\
{\it a. Department of Physics, \\
Purdue University,\\ 
West Lafayette, IN 47907, USA\\
liul@physics.purdue.edu \\

\vspace{10pt}

b. Department of Physics, and \\
Research and Education Center for Natural Sciences\\
Keio University,\\
4-1-1 Hiyoshi, Yokohoma, Kanagawa, 223-8521, Japan\\
nitta@phys-h.keio.ac.jp\\
}}
\end{center}
\vspace{15pt}
{\bf Abstract.} 
The dynamics of the non-Abelian vortex-string, which describes
its low energy oscillations into the target $D=3+1$
spacetime as well as its orientations in the internal space, is 
derived by the approach of nonlinear realization. The resulting 
action correlating these two sectors is found to have an invariant 
synthesis form of the Nambu-Goto-${\bf C}P^{N-1}$ 
model actions. Higher order corrections to the vortex actions are 
presented up to the order of quartic derivatives. General $p$-brane 
dynamics in terms of the internal symmetry breaking is also discussed. 
\vspace{60pt}

\pagebreak

\setcounter{page}{1}
\setcounter{footnote}{0}
\renewcommand{\thefootnote}{\arabic{footnote}}

%\begin{flushleft}
%{\Large I. Introduction}
%\end{flushleft} 
%\vspace{15pt}   
%%%%%%%%%%%%%%%%%%%%%%%%%%%%%%%%%%%%%%%%%%%%%%%%%%%%%%%
\section{Introduction}   
It is well known that topological defects as physical objects can form 
in systems that exhibit the phenomenon of spontaneous symmetry 
breaking. In fact, as early as in the early seventies, the spontaneously 
broken gauge theory was found to possess these soliton-like solutions, 
namely, the Nielsen-Olesen vortex-string \cite{Nielsen:1973cs} and the magnetic monopole 
of 't Hooft and Polyakov \cite{'tHooft:1974qc}, corresponding to Abelian and non-Abelian 
gauge theories, respectively. 

    The vortex-string topological defect, on 
which particular attention has been focused, plays 
important roles in diverse areas of physics, covering 
condensed matter, particle physics as well as cosmology \cite{Vilenkin}. 
In addition, in brane world scenarios, vortices as codimension two branes are deemed 
to be especially useful when considering localization of gauge 
fields by using warped compactifications \cite{Chodos:1999zt}.

   Actually the vortex-string 
topological defect, such as the Nielsen-Olesen vortex line, 
can form if the vacuum manifold $M$ is not simply connected, 
 $\pi_1(M)\neq 0$. The vacuum manifold is $M=U(1)$
in the Abelian Higgs model, 
admitting the Nielsen-Olesen vortex line \cite{Nielsen:1973cs}.
On the other hand, the vortex-strings arising from non-Abelian 
gauge theories have attracted much attention and been extensively 
studied in supersymmetric gauge theories
\cite{Hanany:2003hp}--\cite{Eto:2005yh}, 
see \cite{Tong:2005un} as a review. There, the underlying 
theory is a $U(N_{\rm C} )$ gauge theory with  
$N_{\rm F} (\geq N_{\rm C})$ Higgs fields 
in the fundamental representation with the common $U(1)$ charges.
In the case when $N_{\rm F}  = N_{\rm C}  \equiv N$, the theory is found to have a unique 
vacuum state, i.e. the so called color-flavor locking phase, whose vacuum 
state preserves a global unbroken $SU(N)_{\rm C + F}$ symmetry. Such a system admits the 
solution of Nielsen-Olesen type of vortices, and the solution further breaks 
this locking symmetry to $SU(N - 1) \times U(1)$. Therefore, unlike the Abelian vortices, these non-Abelian vortices have moduli 
(zero modes)
corresponding to spontaneous breaking of the non-Abelian 
symmetry; i.e., besides the usual position moduli they also carry 
orientational moduli ${\bf C}P^{N-1} = SU(N)/[SU(N - 1) \times U(1)]$ 
in the color and flavor internal space.  

Similar non-Abelian vortex-strings appear in QCD with high baryon density 
at low temperature \cite{Balachandran:2005ev} 
where the $SU(3)_{\rm C}$ gauge symmetry is broken by 
diquark condensations 
and is locked with 
the $SU(3)_{\rm F}$ flavor symmetry acting on the light quarks.
Those vortices carry the ${\bf C}P^2$ orientational zero modes 
\cite{Nakano:2007dr,Eto:2009bh} 
and may exist in the core of a neutron star \cite{Sedrakian:2008aya}.

   The dynamics of such non-Abelian vortices, whose underlying theory 
illustrates both the spacetime and internal symmetry breaking, can be 
described by the collective coordinates in such an enlarged moduli space. 
Each moduli parameter provides a massless field for the effective field 
theory on the string world-sheet, which actually corresponds to the 
Nambu-Goldstone scalar associated with the spontaneously broken global symmetry.
 
   As for the spontaneous symmetry breaking, the approach of nonlinear realization
has been demonstrated a natural, economical and elegant framework for treating it.
In fact, this method has been applied to a wide range of physical 
problems most notably in the form of 
nonlinear sigma (NLS) models \cite{GellMann:1960np}, 
supersymmetry \cite{Volkov:1973ix}--\cite{Bagger:1994vj}, 
brane theories \cite{Hughes:1986dn}--\cite{Clark:2005ht}, 
and combination of them. There, the Lagrangian is invariant with respect to the 
transformations of some continuous group $G$, but the ground state is not an invariant 
of $G$ but only of some subgroup $H$. In this context, the resulting phenomenological 
Lagrangian becomes an effective theory at energies far below the scale of spontaneous 
symmetry breaking. Consequently, the effective action can be expressed in terms of the 
dynamics of these Nambu-Goldstone fields.

    In the present context, the formation of the non-Abelian vortex breaks the spacetime and 
internal space symmetries at the same time, i.e., it breaks the target four-dimensional 
Minkowski spacetime to the lower two-dimensional world-sheet spacetime; meanwhile, it also has 
dynamical modes describing its orientations in the special color-flavor locked phase. The interactions 
between these two different sectors would be of interest to the general vortex-string theory and 
is worthy of detail exploration and investigation. Beside, as the string is embedded in a higher 
target spacetime, the terms that characterize this embedding or represent the string rigidity 
may also be supplemented to its actions \cite{Polyakov:1986cs,Curtright:1986vg}.  

    The organization of this paper is as follows. In section \ref{sec:2}, the spacetime 
symmetry breaking of the vortex-string is constructed in terms of the coset structure $ISO(1,3)/[SO(1,3) \times SO(2)]$. 
Then after 
considering the spontaneous breaking of the internal space, the 
kinetic terms associated with the Nambu-Goldstone fields are shown to be described 
by the metric on the internal coset manifold 
${\bf C}P^{N-1} = SU(N)/[SU(N - 1) \times U(1)]$. Therefore, the low energy effective 
actions of the vortex, which illustrate both the spacetime and internal space symmetry 
breakings, are obtained by means of Maurer-Cartan one-forms. As a result, in addition 
to the long wave oscillating modes associated with the translational directions transverse 
to the vortex-string, the string also has oscillating modes corresponding to its 
orientations in the internal space. The effective action between these two sectors is found 
to have a factorized form of the Nambu-Goto-${\bf C}P^{N-1}$ model actions. Besides, the term that describes 
the world-sheet embedding is given by the extrinsic curvature couplings and produces interactions 
containing quartic derivatives. In section  \ref{sec:3}, the general formalism of the $p$-brane dynamics 
corresponding to both spacetime and internal space symmetry breakings is constructed. The 
effective action is then has a synthesis form of the Nambu-Goto-NLS model actions. Section \ref{sec:4} is 
devoted to conclusion and discussion.
In appendix \ref{sec:SF} we give a relation with the Skyrme-Faddeev model 
for $N=1$ (${\bf C}P^1$) with 3+1 dimensional world-volume 
($D=5+1, p=3$).

%\vspace{15pt}
%\begin{flushleft}
%{\Large II. Effective Actions of non-Abelian vortex-strings }
%\end{flushleft}
%\vspace{15pt}
%%%%%%%%%%%%%%%%%%%%%%%%%%%%%%%%%%%%%%%%%%%%%%%%%%%%%%%
\section{Effective Actions of Non-Abelian Vortex-Strings \label{sec:2}} 
Let us consider the non-Abelian vortex-strings described in 
Ref.~\cite{Hanany:2003hp,Auzzi:2003fs}. There, 
the Lagrangian has a $U(N)_{\rm C}$ color symmetry along 
with an $SU(N)_{\rm F}$ flavor symmetry. 
As a result of the overall $U(1)$ gauge symmetry breaking, it ensures the existence 
of the vortex solution for the underlying theory. In addition, the vacuum is 
found to remain a diagonal global color-flavor locking phase with 
$SU(N)_{\rm C + F}$ symmetry.  
Furthermore, the vortex-string solution is found to further break this symmetry 
down to the $SU(N - 1) \times U(1)$ symmetry. Since the diagonal color-flavor symmetry is not broken 
by the VEV of the scalar fields in the bulk, this breaking is physical and has 
nothing to do the Higgs mechanism caused by the gauge transformation. This fact 
therefore leads to the internal orientation moduli space of the string, and the 
presence of these modes makes the string genuinely non-Abelian. The whole moduli 
space of the non-Abelian string then has the form 
${\bf C} \times {\bf C}P^{N-1}$, where ${\bf C}$ is related to the coset 
space $ISO(1,3)/[SO(1,3) \times SO(2)]$, describing the transverse oscillations of the string in the translational 
directions of the world-sheet; while the internal degrees of freedom is given by the 
coset space ${\bf C}P^{N-1} \simeq SU(N)/[SU(N - 1) \times U(1)]$, corresponding to orientational moduli of the string in the internal 
color-flavor $SU(N)$ space. 

     We start first with the dynamics of the string corresponding to the spontaneous 
breaking of the spacetime symmetry in the presence of a vortex-string
\cite{Hughes:1986dn}. Place such a vortex-string along the $x^1$-axis, the 
world-sheet is then parameterized by $\{x^0 ,x^1\}$ in the static gauge, while the target spacetime 
parameterized by the coordinates $x^\mu$ with $\mu=0,1,2,3$.  Therefore, the stability subgroup $H_o$ of the target 
spacetime symmetry breaking is given by the direct product of $SO(1,1) \times SO(2)$, which corresponds to 
the Lorentz boost $SO(1,1)$ symmetry (formed by the generator $M = M^{ab}, a,b = 0,1
$) and the rotational $SO(2)$ invariance 
(formed by the generator $T = M^{23}$) in the $x^2$-$x^3$ plane respectively. Accordingly, the coset 
representative elements $\Omega _o  = G_o /H_o=G_o /[SO(1,1) \times SO(2)]$ 
can be exponentially parameterized as 
\beq
\Omega _o = e^{ix^a P_a } e^{i\phi _i (x)Z_i } e^{iu_i^a (x)K_{ia} }  
\label{1}
\eeq                                                                            in the static gauge; in which $x^a ,\phi _i ,u_i^a$ are the collective coordinates parameterizing the coset 
space $\Omega _o$, and $G_o$ is 
the target $D=3+1$ Poincare group. The broken generators are the automorphism generators $K_1^a  =M^{a2}$ 
and $K_2^a  =M^{a3}$ along with 
the broken spacetime generators $Z_i (Z_i  = Z_1 ,Z_2  = P_2 ,P_3 )$ associated with the 
translational directions transverse to the string. 
We take the convention $\eta ^{ab}  = ( + , - )$ in what follows of the paper.

   The effective action of the string that describes its low energy 
oscillations into the covolume space can be constructed by using the 
Zweibein from the coset structure $\Omega _o ^{ - 1} d\Omega _o$, which has an explicit expansion 
with respect to the $G_o$ generators
\beq
\Omega _o ^{ - 1} d\Omega _o  = i(\omega ^A P_A  + \omega _{z_i } Z_i  + \omega _{k_i }^A K_{iA}  + \omega _T T + \omega _M M) .
\label{2}
\eeq                                                      
It is found that               
\beq
i\omega ^A P_A &=& idx^B P_A (\delta _B ^{\hspace{3pt}A}  + u_{iB} (U^{ - 1} (\cosh \sqrt U  - 1))_{ij} u_j^A ) \non
 && + id\phi _i (U^{ - \frac{1}{2}} \sinh \sqrt U )_{ij} u_j ^A P_A 
\label{3}
\eeq                                                   
where $A,B = 0,1$ and $U_{ij}  = u_i^a u_{ja}$. The capital letters $A,B$ are used to represent the covariant 
spacetime coordinate indices of the string world-sheet, and the lowercase 
letters $a,b$ are used to represent $1+1$ general coordinate indices in what follows. 
Considering $\omega ^A  = dx^b e_{ob} ^{\hspace{3pt}A} $ in static gauge, the Zweibein is therefore found to have the form
\beq
e_{oa} ^{\hspace{3pt}A} =\delta _a ^{\hspace{3pt}A}  + u_{ia} (U^{ - 1} (\cosh \sqrt U  - 1))_{ij} u_j^A  
+ {\partial}_a \phi _i (U^{-1/2} \sinh \sqrt U )_{ij} u _j ^A  .
\label{4}
\eeq                                                       
Under the transformation $g_o \Omega _o  = \Omega '_o h_o$, the coset transforms as $\Omega _o  \to \Omega '_o$, 
and the Maurer-Cartan 
one-forms transform according to
\beq
{\Omega '}_o ^{ - 1} d {\Omega '}_o  = h_o (\Omega _o ^{ - 1} d\Omega _o )h_o ^{ - 1}  + h_o dh_o ^{ - 1} .
\label{5}
\eeq                                                      
Therefore, it can be concluded from Eq.~(\ref{5}) that the covariant coordinate 
differentials have the transformation property 
\beq
i{\omega '}^A P_A  &= & i{dx'}^a {e'}_{oa} ^{\hspace{3pt}A} P_A  \non
  &=& e^{i(bM + \rho T)} idx^b e_{ob} ^{\hspace{3pt}B} P_B e^{ - i(bM + \rho T)}  = idx^b e_{ob} ^{\hspace{3pt}B} L_B ^{\hspace{3pt}A} P_A 
 \label{6}
\eeq                                                                            where $L_B^{\hspace{3pt}A}$ is the representation of the local unbroken $H_o$ symmetry with vector indices,  
and $H_o$ is spanned by the set of generators $\{ M^{ab} ,T \}$. As a result, the transformation of 
Zweibein induced by Eq.~(\ref{6}) has the form
\beq
{e'}_{oa}^{\hspace{3pt}A}  = \frac{{\partial x^b }}{{\partial {x'}^a }}e_{ob}^{\hspace{3pt}^B} L_B^{\hspace{3pt}A}  .
\label{7}
\eeq                                                      
After eliminating the non-dynamic superfluous fields $u_i^a$ with imposing the covariant 
constraint $\omega _z  = 0$ in Eq.~(\ref{2}) by the inverse Higgs mechanism \cite{Ivanov:1975zq},
the metric tensor of the two dimensional world-sheet is found to be 
\beq
g^o_{ab}  = & e_{oa}^{\hspace{3pt}A} e_{ob}^{\hspace{3pt}B} \eta _{AB} % \\ 
  =  (\eta _{ab}  - \partial _a \phi _i \partial _b \phi _i ) .
\label{8}
\eeq

%%%%%%%%%%%%%%%%%%%%%%%%%%%%%
   On the other hand, the formation of the vortex topological defect breaks 
the color-flavor locking symmetry $SU(N)_{\rm C+F}$ to the stability subgroup $SU(N - 1) \times U(1)$, and this embedding 
is rather a dynamical process. Therefore, the dynamics of the vortex-string 
corresponding to the internal orientional modes is described by these collective 
coordinates of the internal coset space $SU(N)/[SU(N - 1) \times U(1)]$. Consider the Lie algebra of $SU(N)$ group, 
which has $N^2  - 1$ generators. In the fundamental representation they have the form
\beq
(T_{a'b'}^1 )_{c'd'}  &=& (\delta _{a'c'} \delta _{b'd'}  + \delta _{b'c'} \delta _{a'd'} ) ,\non
(T_{a'b'}^2 )_{c'd'}  &=&  - i(\delta _{a'c'} \delta _{b'd'}  - \delta _{b'c'} \delta _{a'd'} ) 
\label{9}
\eeq                                                      
where $a',b',c',d' = 1,2,3,\cdots,N$ and $a' < b'$. 
The diagonal generators can be written as  
\beq
(T_{a'}^3 )_{c'd'}   
  = \left\{ {\begin{array}{*{20}cc}
   {\delta _{c'd'} \sqrt {\frac{2}{{a'(a' - 1)}}} ,}  & {c' < a'}  \\
   { - \delta _{c'd'} \sqrt {\frac{{2(a' - 1)}}{{a'}},} }  & {c' = a'}  \\
   {0,}  &  {c' > a'}  \\
%\end{array}\begin{array}{*{20}c}
%   {}  \\
%   {}  \\
%   {}  \\
%\end{array}\begin{array}{*{20}c}
\end{array}} \right. 
\label{10}
\eeq                                                      
where $2 \le a' \le N$. These matrices have the normalization condition  
\beq
{\rm Tr}(T_{A'} T_{B'} ) = 2\delta _{A'B'}   
\label{11}
\eeq                                                      
where $A',B'$ are from $1$ to $N^2 - 1$. Likewise, the generators of $SU(N - 1)$ group can be taken as
\beq
(T_{a'b'}^1 )_{c'd'}  &=& (\delta _{a'c'} \delta _{b'd'}  + \delta _{b'c'} \delta _{a'd'} ), \non
(T_{a'b'}^2 )_{c'd'}  &=&  - i(\delta _{a'c'} \delta _{b'd'}  - \delta _{b'c'} \delta _{a'd'} ) 
\label{12}
\eeq                                                        
with $a',b',c',d' = 1,2,3,\cdots,N - 1$ and $a' < b'$, and the generator of the $U(1)$ group has the diagonal 
traceless form                                                                  \beq
(T_N^3 )_{c'd'}  
  = \left( {\begin{array}{*{20}c}
   {\sqrt {2/[N(N - 1)]} } & {} & {} & {} & {}  \\
   {} & \ddots & {} & {} & {}  \\
   {} & {} & \ddots & {} & {}  \\
   {} & {} & {} & {\sqrt {2/[N(N - 1)]} } & {}  \\
   {} & {} & {} & {} & { - \sqrt {2(N - 1)/N} }  \\
\end{array}} \right) \quad
\label{13}
\eeq                           
i.e. 
\beq
(T_N^3 )_{c'd'}   
  = \left\{ {\begin{array}{*{20}cc}
   {\delta _{c'd'} \sqrt {\frac{2}{{N(N - 1)}}} ,}  & {c' < N}  \\
   { - \delta _{c'd'} \sqrt {\frac{{2(N - 1)}}{N},} } &  {c' = N}  \\
\end{array}} \right. 
\label{14}
\eeq                           
In addition, the broken generators associated with the coset 
$SU(N)/[SU(N - 1) \times U(1)]$ are given 
by $N-1$ generators of the first type
\beq
(T_{a'N}^1 )_{c'd'}  = (\delta _{a'c'} \delta _{Nd'}  + \delta _{Nc'} \delta _{a'd'} ) 
\label{15}
\eeq     
along with the other $N-1$ generators of the second type
\beq
(T_{a'N}^2 )_{c'd'}  =  - i(\delta _{a'c'} \delta _{Nd'}  - \delta _{Nc'} \delta _{a'd'} )  
\label{16}
\eeq      
in which $a' = 1,2,3,\cdots,N - 1$, and $c',d' = 1,2,3,\cdots,N$. Therefore we have total 
$N^2  - 1 - ((N - 1)^2  - 1) + 1) = 2N - 2 $ generators corresponding to 
the coset space, which have explicit forms  
\beq
&& T_{}^{1a'}  = T_{a'N}^1  = \left( {\begin{array}{*{20}c}
   0 & {\cdots} & 0 & 1  \\
   \vdots & \ddots & {} & 0  \\
   0 & {} & \ddots & \vdots  \\
   1 & 0 & {\cdots} & 0  \\
\end{array}} \right), 
% \left( {\begin{array}{*{20}c}
%   0 & {\cdots} & 0 & 0  \\
%   \vdots & \ddots & {} & 1  \\
%   0 & {} & \ddots & \vdots  \\
%   0 & 1 & {\cdots} & 0  \\
%\end{array}} \right),
\cdots,
 \left( {\begin{array}{*{20}c}
   0 & {\cdots} & 0 & 0  \\
   \vdots & \ddots & {} & \vdots  \\
   0 & {} & \ddots & 1  \\
   0 & {\cdots} & 1 & 0  \\
\end{array}} \right) ,\non
&& 
T_{}^{2a'}  = T_{a'N}^2  
  = \left( {\begin{array}{*{20}c}
   0 & {\cdots} & 0 & { - i}  \\
   \vdots & \ddots & {} & 0  \\
   0 & {} & \ddots & \vdots  \\
   i & 0 & {\cdots} & 0  \\
\end{array}} \right) , 
% \left( {\begin{array}{*{20}c}
%   0 & {\cdots} & 0 & 0  \\
%   \vdots & \ddots & {} & { - i}  \\
%   0 & {} & \ddots & \vdots  \\
%   0 & i & {\cdots} & 0  \\
%\end{array}} \right) ,
\cdots ,
 \left( {\begin{array}{*{20}c}
   0 & {\cdots} & 0 & 0  \\
   \vdots & \ddots & {} & \vdots  \\
   0 & {} & \ddots & { - i}  \\
   0 & {\cdots} & i & 0  
\end{array}} \right).  
\label{17}
\eeq       
                  
Therefore, the coset representative elements $\Omega _{\cal I}  = G_{\cal I} /H_{\cal I} =SU(N)/[SU(N - 1) \times U(1)]$ 
with respect to the stability group $H_{\cal I}$ can be exponentially parameterized as \beq
\Omega _{\cal I}  = e^{i\left(\phi ^{1a'} T^{1a'}  + \phi ^{2a'} T^{2a'} \right)}.  
\label{18}
\eeq                     
The parameters $\phi ^{1a'}$ and $\phi ^{2a'}$ are the Nambu-Goldstone (NG) fields corresponding 
to the spontaneous breaking of the full $SU(N)_{\rm C+F}$ group. They transform nonlinearly under 
the left action of the general group elements $g_{\cal I}$ on the coset representative 
itself, i.e. $g_{\cal I} \Omega _{\cal I}  = \Omega '_{\cal I} h_{\cal I} $, and the resulting inhomogeneous terms of the NG transformations 
directly signal the breakdown of the $SU(N)_{\rm C+F}$ symmetry. The effective action of the string 
that describes its orientional modes in the internal space can be derived by using 
the metric tensor on the coset manifold, which is constructed from the Maurer-Cartan 
one-forms $\Omega _{\cal I} ^{ - 1} d\Omega _{\cal I} $. In which the coset elements can be rewritten as
\beq
\Omega _{\cal I}  = e^{i(\phi ^{1a'} T^{1a'}  + \phi ^{2a'} T^{2a'} )} = e^{i(\Phi ^{i'} T^{i'}  + \Phi ^{*i'} T^{i'\dagger} )}  
\label{19}
\eeq                  
where $\Phi ^{i'}  = \phi ^{1i'}  + i\phi ^{2i'}$, $\Phi ^{*i'}  = \phi ^{1i'}  - i\phi ^{2i'} $ 
with $\phi$ and $\Phi$ real and complex scalar fields, respectively,  and 
\beq
&& T_{}^{i'}  
  = \left( {\begin{array}{*{20}c}
   0 & {\cdots} & 0 & 0  \\
   \vdots &  \ddots & {} & 0  \\
   0 & {} &  \ddots & \vdots \\
   1 & 0 & {\cdots} & 0  \\
\end{array}} \right) , 
 \left( {\begin{array}{*{20}c}
   0 & {\cdots} & 0 & 0  \\
   \vdots &  \ddots & {} & 0  \\
   0 & {} &  \ddots & \vdots  \\
   0 & 1 & {\cdots} & 0  \\
\end{array}} \right) ,\cdots,
 \left( {\begin{array}{*{20}c}
   0 & {\cdots} & 0 & 0  \\
   \vdots &  \ddots & {} & \vdots  \\
   0 & {} &  \ddots & 0  \\
   0 & {\cdots} & 1 & 0  \\
\end{array}} \right), \non
&& T_{}^{i' \dagger }   
  = \left( {\begin{array}{*{20}c}
   0 & {\cdots} & 0 & 1  \\
   \vdots &  \ddots & {} & 0  \\
   0 & {} &  \ddots & \vdots   \\
   0 & 0 & {\cdots} & 0  \\
\end{array}} \right) , 
 \left( {\begin{array}{*{20}c}
   0 & {\cdots} & 0 & 0  \\
   \vdots &  \ddots & {} & 1  \\
   0 & {} &  \ddots & \vdots   \\
   0 & 0 & {\cdots} & 0  \\
\end{array}} \right) ,\cdots,
 \left( {\begin{array}{*{20}c}
   0 & {\cdots} & 0 & 0  \\
   \vdots &  \ddots & {} & \vdots   \\
   0 & {} &  \ddots & 1  \\
   0 & {\cdots} & 0 & 0  \\
\end{array}} \right), \non
\label{20}
\eeq 
where $i' = 1,2,3,\cdots,N - 1$. Accordingly, these matrices have explicit multiplication relations as follows
\beq
&& T_{}^{i'} T_{}^{j'} = 0;   T_{}^{i' \dagger } T_{}^{j' \dagger } = 0; T_{}^{i'} T_{}^{j' \dagger}  = T\delta ^{i'j'};  \non
&& T_{}^{i' \dagger } T_{}^{j'}  = M^{i'j'}; T_{}^{i'} T  = 0; T_{}^{j' \dagger }  T  = T_{}^{j' \dagger };  \non
&& T T_{}^{i'} = T_{}^{i'}; T T_{}^{i' \dagger } = 0; T T  = T ; \non
&&T_{}^{i'} M^{j'k'}  = T^{k'} \delta ^{i'j'}; T_{}^{i' \dagger }M^{j'k'}= 0 ; M^{i'j'} T_{}^{k'}= 0; \non
&& M^{k'j'} T_{}^{i' \dagger } = T^{k' \dagger } \delta ^{i'j'}; M^{i'j'} T = 0; T M^{i'j'}= 0; 
\label{21}
\eeq
where                                                     
\beq
T = \left( {\begin{array}{*{20}c}
   0 & {\cdots} & 0 & 0  \\
   \vdots & \ddots  & {} & 0  \\
   0 & {} & \ddots  & \vdots   \\
   0 & 0 & {\cdots} & 1  \\
\end{array}} \right) 
\label{22}
\eeq
and 
\beq
M^{11}  = \left( {\begin{array}{*{20}c}
   1 & {\cdots} & 0 & 0  \\
   0 & \ddots  & {} & 0  \\
   \vdots & {} & \ddots  & \vdots  \\
   0 & 0 & {\cdots} & 0  
\end{array}} \right) , \quad 
 M^{12}  = \left( {\begin{array}{*{20}c}
   0 & 1 & {\cdots} & 0  \\
   0 & \ddots  & {} & 0  \\
   \vdots & {} & \ddots  & \vdots  \\
   0 & 0 & {\cdots} & 0  \\
\end{array}} \right)  
,\cdots
\eeq
\beq
 M^{12}  = \left( {\begin{array}{*{20}c}
   0 & 0 & {\cdots} & 0  \\
   1 & \ddots  & {} & 0  \\
   {} & {} & \ddots  & \vdots  \\
   0 & 0 & {\cdots} & 0  \\
\end{array}} \right) , \quad  
 M^{21}  = \left( {\begin{array}{*{20}c}
   0 & 0 & {\cdots} & 0  \\
   0 & 1 & {\cdots} & 0  \\
   \vdots & {} & \ddots  & \vdots  \\
   0 & 0 & {\cdots} & 0  \\
\end{array}} \right) ,\cdots .
\label{23}
\eeq
After normalizing the Nambu-Goldstone fields 
in the coset representative element,                                            \beq
\Omega _{\cal I}  =  e^{if(|\Phi |)(\frac{{\Phi ^{i'} }}{{|\Phi |}}T^{i'}  + \frac{{\Phi ^{*i'} }}{{|\Phi |}}T^{i'} )}  
\label{24}
\eeq                                                                            
in which $f(|\Phi|)$ is the normalization function determined below, and 
the norm is defined as $|\Phi | = \sqrt {\Phi ^{i'} \Phi ^{*i'} }$, 
we can explicitly expand it with respect to the $G_{\cal I}$ generators, i.e.   
\beq
\Omega _{\cal I} &= & 1 + (\cos (f(|\Phi |)) - 1)T + \frac{{\cos (f(|\Phi |)) - 1}}{{|\Phi |^2 }}\Phi ^{*i'} \Phi ^{j'} M^{i'j'}  \non
  && + i\frac{{\sin (f(|\Phi |))}}{{|\Phi |}}(\Phi ^{*i'} T^{i' \dagger }  + \Phi ^{i'} T^{i'} )
 \label{25}
\eeq   
likewise
\beq
\Omega _{\cal I}^{ - 1}  &= & 1 + (\cos (f(|\Phi |)) - 1)T + \frac{{\cos (f(|\Phi |)) - 1}}{{|\Phi |^2 }}\Phi ^{*i'} \Phi ^{j'} M^{i'j'}  \non 
 && - i\frac{{\sin (f(|\Phi |))}}{{|\Phi |}}(\Phi ^{*i'} T^{i' \dagger }  + \Phi ^{i'} T^{i'} ) .                               
 \label{26}
\eeq         
The Maurer-Cartan one-forms is then found to be                                
\beq
\Omega _{\cal I}^{ - 1} d\Omega _{\cal I}  = i(A^{a'} T^{a'}  + B^{a'} T^{a' \dagger }  + CT + D^{a'b'} M^{a'b'} ) 
\label{27}
\eeq     
where         
\beq
A^{a'}  &= & \cos (f(|\Phi |)) d\left(\sin (f(|\Phi |))\frac{{\Phi ^{a'} }}{{|\Phi |}}\right) \non
 && - \frac{{\sin (f(|\Phi |))}}{{|\Phi |}}
 d\left[\left\{\cos (f(|\Phi |)) - 1\right\}\frac{{\Phi ^{*j'} \Phi ^{a'} }}
        {{|\Phi |^2 }}\right]\Phi ^{j'}  \non
 &= & df\frac{{\Phi ^{a'} }}{{|\Phi |}} 
 - (\cos f - 1)\sin f\frac{{\Phi ^{a'} }}{{|\Phi |}}
  d\left(\frac{{\Phi ^{*j'} }}{{|\Phi |}}\right)
  \frac{{\Phi ^{j'} }}{{|\Phi |}} 
  + \sin f d\left(\frac{{\Phi ^{a'} }}{{|\Phi |}}\right); \non
 B^{a'} &= & df\frac{{\Phi ^{*a'} }}{{|\Phi |}} 
  - (\cos f - 1)\sin f\frac{{\Phi ^{*a'} }}{{|\Phi |}}
    d\left(\frac{{\Phi ^{j'} }}{{|\Phi |}}\right)
    \frac{{\Phi ^{*j'} }}{{|\Phi |}} 
 + \sin fd\left(\frac{{\Phi ^{*a'} }}{{|\Phi |}}\right); \non 
 C  &= & - i\sin f\frac{{\Phi ^{a'} }}{{|\Phi |}}
     d\left(\sin f\frac{{\Phi ^{*a'} }}{{|\Phi |}}\right) 
  + \cos f\sin f \;df ;\non
 D^{a'b'} &= & - i\sin f\frac{{\Phi ^{*a'} }}{{|\Phi |}}
     d\left(\sin f\frac{{\Phi ^{b'} }}{{|\Phi |}}\right) 
  - id\left[(\cos f - 1)\frac{\Phi ^{*a'} \Phi ^{b'}}{{|\Phi |^2 }} \right] 
   \non 
  && - i(\cos f - 1)\frac{{\Phi ^{*a'} \Phi ^{c'} }}{{|\Phi |^2 }}
   d\left[(\cos f - 1)\frac{{\Phi ^{*c'} \Phi ^{b'} }}{{|\Phi |^2 }}\right]. 
 \label{28}
\eeq                   
Under the action $g_{\cal I} \Omega _{\cal I}  = \Omega '_{\cal I} h_{\cal I}$, one can find that the Maurer-Cartan 
one-forms transform according to 
\beq
{\Omega '}_{\cal I} ^{ - 1} d {\Omega '}_{\cal I}  = h_{\cal I} (\Omega _{\cal I} ^{ - 1} d\Omega _{\cal I} )h_{\cal I} ^{ - 1}  + h_{\cal I} dh_{\cal I} ^{ - 1}  .
\label{29}
\eeq      
Among them, the one-forms associated with the broken generators 
$T^{a'}$ and $T^{a' \dagger }$ transform 
covariantly, and the one-forms associated with the unbroken generators $T$ and $M^{a'b'}$ transform 
inhomogeneously.

    In addition, after rewriting the one-forms $A^{a'}$ as  
\beq
&&A^{a'}  =  \frac{{df}}{{d|\Phi |}}\frac{{\Phi ^{a'} \Phi ^{*i'} d\Phi ^{i'} }}{{2|\Phi |^2 }} + (\cos f - 1)\sin f\frac{{\Phi ^{a'} }}{{2|\Phi |^2 }}\frac{{\Phi ^{*i'} d\Phi ^{i'} }}{{|\Phi |}} \non
 & & \quad\quad
 + \sin f \left( - \frac{{\Phi ^{*i'} d\Phi ^{i'} \Phi ^{a'} }}{{2|\Phi |^3 }} + \frac{{d\Phi ^{a'} }}{{|\Phi |}}\right)  
  + \frac{{df}}{{d|\Phi |}}\frac{{\Phi ^{a'} d\Phi ^{*i'} \Phi ^{i'} }}{{2|\Phi |^2 }} \non 
 &&\quad\quad 
 - (\cos f - 1)\sin f\frac{{\Phi ^{a'} }}{{2|\Phi |^2 }}\frac{{\Phi ^{i'} d\Phi ^{*i'} }}{{|\Phi |}} 
  - \sin f\frac{{d\Phi ^{*i'} \Phi ^{i'} \Phi ^{a'} }}{{2|\Phi |^3 }} 
\label{30}
\eeq                   
one can obviously note that the two covariant parts, 
which correspond to $d\Phi$ and $d\Phi ^*$ terms respectively, transform independently under 
Eq.~(\ref{29}). 
Imposing the covariant condition on this one-forms by setting the 
term related to $d\Phi ^*$ to zero gives us
\beq
 \left(\frac{{df}}{{d|\Phi |}}\frac{1}{{2|\Phi |^2 }} - \cos f\sin f\frac{1}{{2|\Phi |^3 }}\right)\Phi ^{a'} d\Phi ^{*i'} \Phi ^{i'} = 0 .
\label{31}
\eeq
Then the normalization function can be secured as 
\beq
 df/d|\Phi | = \cos f\sin f\frac{1}{{|\Phi |}} ,
\label{32}
\eeq
i.e.
\beq
f = \arctan (|\Phi |) .
\label{33}
\eeq
Therefore, $A^{a'}$ becomes 
\beq
A^{a'} &= & \frac{{d\Phi ^{i'} \Phi ^{*i'} }}{{1 + |\Phi |^2 }}\frac{{\Phi ^{a'} }}{{2|\Phi |^2 }} +  
 \sin f\frac{1}{{|\Phi |}}d\Phi ^{a'}  - \sin f\Phi ^{*i'} \frac{{d\Phi ^{i'} }}{{|\Phi |^3 }}\Phi ^{a'}   \non 
 & & + \cos f\sin f \Phi ^{*i'} \frac{{d\Phi ^{i'} }}{{2|\Phi |^3 }}\Phi ^{a'} 
  \non 
 &= & \frac{1}{{\sqrt {1 + |\Phi |^2 } }} 
 \left(d\Phi ^{a'}  - \frac{1}{{|\Phi |^2 }}\Phi ^{*i'} d\Phi ^{i'} \Phi ^{a'}  + \frac{{d\Phi ^{i'} \Phi ^{*i'} }}{{\sqrt {1 + |\Phi |^2 } }}\frac{{\Phi ^{a'} }}{{2|\Phi |^2 }} \right.\non
&&\left. +  \frac{1}{{\sqrt {1 + |\Phi |^2 } }} 
 \Phi ^{*i'} \frac{{d\Phi ^{i'} }}{{2|\Phi |^2 }}\Phi ^{a'}  \right)\non 
&= & \frac{1}{{\sqrt {1 + |\Phi |^2 } }}
  \left[d\Phi ^{a'}  + \left( - \frac{1}{{|\Phi |^2 }}
  + \frac{1}{{\sqrt {1 + |\Phi |^2 } }}\frac{1}{{|\Phi |^2 }}\right) 
    \Phi ^{*i'} \Phi ^{a'} d\Phi ^{i'} \right] 
\label{34}
\eeq                   
where $A^{a'}$ is the covariant coordinate differential, and $d\Phi ^{i'}$ is the general coordinate 
differential on the coset manifold.
The vielbein can be derived accordingly 
\beq
A^{a'}  &= & d\Phi ^{i'} e_{{\cal I}i'} ^{\hspace{3pt}a'}  \non 
&= & d\Phi ^{i'} \frac{1}{{\sqrt {1 + |\Phi |^2 } }}
  \left[\delta _{i'} ^{\hspace{3pt}a'}  
   + \left( - \frac{1}{{|\Phi |^2 }}
    + \frac{1}{{\sqrt {1 + |\Phi |^2 } }}\frac{1}{{|\Phi |^2 }}\right) 
    \Phi ^{*i'} \Phi ^{a'} \right].
\label{35}
\eeq                   
Thus the vielbein becomes
\beq
 e_{{\cal I}i'}^{\hspace{3pt}{a'}}= \frac{1}{{\sqrt {1 + |\Phi |^2 } }}
 \left[\delta _{i'} ^{\hspace{3pt}a'}  + \left( - \frac{1}{{|\Phi |^2 }} 
  + \frac{1}{{\sqrt {1 + |\Phi |^2 } }}\frac{1}{{|\Phi |^2 }}\right)  \Phi ^{*i'} \Phi ^{a'} \right] .
\label{36}
\eeq
The $SU(N - 1) \times U(1)$ invariant interval is therefore given by
\beq
ds_{\cal I} ^2  &= & A^{a'} A^{*a'}  = e_{{\cal I}i'} ^{\hspace{3pt}a'} e_{{\cal I}j'}^{*\hspace{3pt}b'} d\Phi ^{i'} d\Phi ^{*j'} \delta _{b'}^{a'}  \non 
  &= & g^{\cal I}_{ij^*} d\Phi ^i d\Phi ^{*j}  
\label{37}
\eeq                   
where the metric $g^{\cal I}_{i'j'^*}$ is 
\beq
g^{\cal I}_{i'j'^*}  = e_{{\cal I}i'} ^{\hspace{3pt}a'} e_{{\cal I}j'}^{*\hspace{3pt}a'}.  
\label{38}
\eeq
Considering Eq.~(\ref{36}), 
the invariant interval of the internal space becomes
\beq
ds_{\cal I} ^2  = g^{\cal I}_{i'j'^*} d\Phi ^{i'} d\Phi ^{*j'} 
 = \frac{1}{{1 + |\Phi |^2 }}\left[d\Phi ^{i'} d\Phi ^{*i'}  
    - \frac{1}{{1 + |\Phi |^2 }} 
\Phi ^{j'} \Phi ^{*i'} d\Phi ^{i'} d\Phi ^{*j'} \right], 
\eeq
where the metric $g^{\cal I}_{i'j'^*}$ can be obtained, %from Eq.(38)
to yield
\beq
g^{\cal I}_{i'j'^*}= \frac{1}{{1 + |\Phi |^2 }}\left[\delta _{i'j'}  + \frac{1}{{\left| \Phi  \right|^2 }}\left(\frac{1}{{1 + |\Phi |^2 }} - 1\right)\Phi ^{j'} \Phi ^{*i'} \right] .\label{39}
%\label{(50)}
\eeq
We thus have obtained the Fubini-Study metric on 
the complex projective space ${\bf C}P^{N-1}$ 
as expected.

\medskip
Now we are ready to construct the low energy 
effective action of the vortex-string. Under the general coordinate 
transformations of Eq.~(\ref{7}), by using Eq.~(\ref{8}) one can 
find the following invariance 
\beq
d^2 {x'}\sqrt {\left| {\det {g^o{}'}} \right|}  
 &= & d^2 {x'}\det {e'}_{ob} ^{\hspace{3pt}A}  
  = d^2 x\det \left| {\frac{{\partial {x'}^b }}{{\partial x^a }}} 
    \right|\det \left| {\frac{{\partial x^b }}{{\partial {x'}^a }}} 
    \right|\det e_{ob} ^{\hspace{3pt}B} \det L_B ^{\hspace{3pt}A}  \non
 &= & d^2 x\det e_{ob} ^{\hspace{3pt}B}  = d^2 x\sqrt {|\det g^o |} 
\label{40}
\eeq                   
along with the invariant internal space interval  
\beq
ds_{\cal I} ^2  = g^{\cal I}_{i'j'^*} d\Phi ^{i'} d\Phi ^{*j'}  
=  ds_{\cal I} '^2  = g^{\cal I}{}'_{i'j'^*}{}d\Phi '^{i'} d\Phi '^{*j'}  .
\label{41}
\eeq
Hence, 
\beq
d^2 x\sqrt {|\det g^o |} g^{\cal I}_{i'j'^*} \partial _a \Phi ^{i'} \partial ^a \Phi ^{*j'}  
  = d^2 x'\sqrt {|\det g^o{}'|} 
 g^{\cal I}{}'_{i'j'^*} \partial '_a \Phi '^{i'} \partial '^a \Phi '^{*j'} 
\label{42}
\eeq
is invariant under both the spacetime and internal space transformations.

Considering Eq.~(\ref{8}) the action of the string described by the 
Nambu-Goldstone oscillation modes corresponding to its low energy oscillations into 
the covolume space takes the form
\beq
S_o  &= & - T\int {d^2 x} \sqrt {\left| {\det g^o } \right|}  =  - T\int {d^2 x\sqrt {|\det (\eta _{ab}  - \partial _a \phi _i \partial _b \phi _i )|} }  \non 
  &= & - T\int {d^2 x\sqrt {\det (\delta _{ij}  - \partial _a \phi _i \partial ^a \phi _j )} } 
\label{43}
\eeq                   
where $T$ stands for string tension, and it is obviously $SO(1,1) \times SO(2)$ invariant. 
This part is the Nambu-Goto action \cite{Nambu:1974zg} 
which is sufficient for describing the dynamics of an Abelian vortex-string.

   In addition to the Nambu-Goto action (\ref{43}) for the broken translational 
zero mode, there exist the Nambu-Goldstone modes for the internal symmetry breaking 
in the case of non-Abelian vortex-strings. The effective action that describes 
the dynamics of the string in both the spacetime 
and internal space breakings is then given by 
\beq
S_{\cal I}  &= & T_{\cal I} \int {d^2 x} \sqrt {|\det g^o |} 
 g^{\cal I}_{i'j'^*} \partial _a \Phi ^{i'} \partial ^a \Phi ^{*j'}  \non 
  &= & T_{\cal I} \int {d^2 x} \sqrt {\det (\delta _{ij}  - \partial _a \phi _i \partial ^a \phi _j )} \non
 && \quad \times \frac{1}{{1 + |\Phi |^2 }} 
 \left[\partial _a \Phi ^{i'} \partial ^a \Phi ^{*i'}  -  
  \frac{1}{{1 + |\Phi |^2 }} \Phi ^{j'} \Phi ^{*i'} 
  \partial _a \Phi ^{i'} \partial ^a \Phi ^{*j'} \right] \non
  &= & T_{\cal I} \int {d^2 x} \{ \partial _a \Phi ^{i'} \partial ^a \Phi ^{*i'}  - 
\frac{1}{2}\partial _a \phi ^i \partial ^a \phi ^i \partial _b \Phi ^{i'} 
\partial ^b \Phi ^{*i'} \non
&& +{\cal O}(\Phi \Phi^* \partial \Phi \partial \Phi^*)+ 
{\cal O} (\partial \Phi \partial \Phi^* (\partial \phi)^4)\},
\label{44}
\eeq            
where $T_{\cal I}$ is the coupling 
between the space-time and the internal zero modes. 
Therefore one can find that the effective action has a synthesis 
form of the Nambu-Goto-${\bf C}P^{N-1}$ model actions $S_o+S_{\cal I}$. 
Note that the moduli space is the direct product 
${\bf C} \times {\bf C}P^{N-1}$ of the translational 
and the internal zero modes 
and consequently the action is decomposed into them 
at the order of the quadratic derivatives, 
but there exist the interaction terms 
at the quartic (and higher) derivatives.

\if0 %%%%%%%%%%
     Specifically, when one takes $N=2$, such as the vortex example considered in 
Ref.~\cite{Auzzi:2003fs}, the vortex topological defect solution breaks the exact color-flavor 
locking phase $SU(2)_{C + F}$ symmetry down to the stability group $U(1)$, i.e.
\beq
SU(2)_{\rm C + F}  \to U(1) .
\label{45}
\eeq
Therefore the internal orientional modes are parameterized by the collective 
coordinates of the internal coset space $SU(2)/U(1)$. 
The dynamics of the vortex thus has an explicit form  
\beq
S_I  = & T_I \int {d^2 x} \sqrt {|\det g^o |}  \frac{1}{{1 + \Phi {\Phi^*}}}
 \left[\partial _a \Phi \partial ^a \Phi ^*  
  - \frac{1}{{1 + \Phi {\Phi^ *}}}\Phi \Phi ^* \partial _a \Phi \partial ^a \Phi ^* \right] \non
  = & T_I \int {d^2 x\sqrt {\det (\delta _{ij}  - \partial _a \phi _i \partial ^a \phi _j )} } \partial _a \Phi \partial ^a \Phi ^* \frac{1}{{(1 + \Phi {\Phi^ *})^2 }} \\ 
  = & T_I \int {d^2 x\{ } \partial _a \Phi \partial ^a \Phi ^*  - \frac{1}{2}\partial _a \phi ^i 
\partial ^a \phi ^i \partial _b \Phi \partial ^b \Phi ^*  - 2\partial _a \Phi \partial ^a \Phi ^* 
\Phi \Phi ^*  + \cdots\} 
 \label{46}
\eeq                   
\fi %%%%%%%%%%%%%%%
                            
Typically, the coherence length and penetration depth 
of the vortex are characterized by the Compton wave lengthes of massive Higgs and gauge bosons respectively.
Here, we have assumed that the width or transverse size of a vortex-string
is much smaller than the energy scale which we consider.
On the other hand, one may consider 
the correction to the action by taking account the effect of width
%the geometrical aspect of the structure 
of the vortex. 
%As the vortex-string embedded in higher dimensional spacetime, 
In order to do so, it
is necessary to add the effective action terms 
that characterizes this embedding 
of the world-sheet in the target spacetime \cite{Polyakov:1986cs,Curtright:1986vg}. The interactions that 
describes the stiffness or rigidity of the string can be shown as 
\beq
S' = k_0 \int {d^2 x} \sqrt {|\det (\eta _{ab}  - \partial _a \phi ^i \partial _b \phi ^i )|} K_{ab}^i K^{iab}  
\label{47}
\eeq
in which the coupling constant $k_0$ stands for the rigidity 
or stiffness, and the extrinsic curvatures $K_{ab}^i$ has the form $K_{ab}^i  = n_\mu ^i \partial _a \partial _b x^\mu$ with 
the normalization condition $n_\mu ^i n^{j\mu }  = \delta ^{ij}$ for the unit 
norm vectors $n_\mu ^i$. As a result, these extrinsic curvatures terms produce interactions 
containing quartic derivatives in the strings actions. 
In general for a $p$-brane  
the intrinsic scalar curvature $R$ on its world-volume 
can be written as $R = (K^{ia}_a)^2 - K^{ia}_b K^{ib}_a$.
In the case of the string, this is a total derivative term 
because of the Euler's theorem, 
and so we do not need it \cite{Polyakov:1986cs}.
Up to the same order 
corrections, the string actions can also be supplemented with the following 
but not scale invariant interactions 
\beq
S_1  = k_1 \int {d^2 x} \sqrt {|\det g^o |} (\partial _a x^\mu  \partial ^a x_\mu  )^2  + k_2 \int {d^2 x} \sqrt {|\det g^o |} \partial _a x^\mu  \partial ^a x_\nu  \partial _b x_\mu  \partial ^b x^\nu .
\label{48}
\eeq
Likewise, we have the following quartic coupling terms for the internal moduli 
\beq
&& S_2  =  k'_1 \int {d^2 x} \sqrt {|\det g^o |} (g^{\cal I}_{i'j'^*} \partial _a \Phi ^{i'} \partial ^a \Phi ^{*j'})^2   \non
  && \quad \;\; 
  + k'_2 \int {d^2 x} \sqrt {|\det g^o |} g^{\cal I}_{i'j'^*} \partial _a \Phi ^{i'} \partial ^b \Phi ^{*j'} g^{\cal I}_{k'l'^*} \partial _b \Phi ^{k'} \partial ^a \Phi ^{*l'} \non
  && \quad \;\; 
  + k'_3 \int {d^2 x} \sqrt {|\det g^o |} g^{\cal I}_{i'j'^*} \partial _a \Phi ^{i'} \partial ^b \Phi ^{*j'} g^{\cal I}_{k'l'^*} \partial _b \Phi ^{*l'} \partial ^a \Phi ^{k'} 
\label{49}
\eeq                   
where the metric $g^{\cal I}_{i'j'^*}$ is given in Eq.~(\ref{39}).\footnote{
In more general, the following generally covariant terms are also invariant 
under the holomorphic isometry $G_{\cal I}$ for 
the K\"ahler internal target spaces:
$\int {d^2 x} \sqrt {|\det g^o |} R^{\cal I}$,  
$\int {d^2 x} \sqrt {|\det g^o |} R^{\cal I} g^{\cal I}_{i'j'^*} \partial _a \Phi ^{i'} \partial ^a \Phi ^{*j'}$, 
$\int {d^2 x}\sqrt {|\det g^o |} R^{\cal I}_{i'j'^*} \partial _a \Phi ^{i'} \partial ^a \Phi ^{*j'}$, 
 $\int {d^2 x}\sqrt {|\det g^o |} R^{\cal I}_{i'j'^*k'l'^*} 
  \partial _a \Phi ^{i'} \partial ^a \Phi ^{*j'} 
  \partial _b \Phi ^{k'} \partial ^b \Phi ^{*l'}$ 
and so on,
where $R^{\cal I}$, $R^{\cal I}_{i'j'^*}$ and $R^{\cal I}_{i'j'^*k'l'^*}$ are the scalar curvature, 
the Ricci-form and the Riemann curvature tensor of 
the target K\"ahler manifold, respectively. 
This is because (holomorphic) isometries are subgroups 
of the (holomorphic) general coordinate 
transformation.  
However those terms are not independent from the terms in Eq.~(\ref{49}) 
in the case of ${\bf C}P^{N-1}$ which we are concerned with 
because of the identities
$R_{\cal I} =$ const., 
$g^{\cal I}_{i'j'^*} = N
R^{\cal I}_{i'j'^*}$ (Einstein), 
and $R^{\cal I}_{i'j'^*k'l'^*} \sim 
(g^{\cal I}_{i'j'^*} g^{\cal I}_{k'l'^*}  + g^{\cal I}_{i'l'^*} g^{\cal I}_{k'j'^*} )$ (symmetric space), see e.g. \cite{Higashijima:2001fp}. 
The so-called K\"ahler normal coordinates should be useful for 
the expansion of geometric quantities 
in the internal K\"ahler manifold \cite{Higashijima:2000wz}.
The above terms are needed in general for arbitrary K\"ahler target spaces.
}
%, which yields
%$$
%g_{Ii'j'}= \frac{1}{{1 + |\Phi |^2 }}[\delta _{i'j'}  + \frac{1}{{\left| \Phi  %\right|^2 }}(\frac{1}{{1 + |\Phi |^2 }} - 1)\Phi ^{j'} \Phi ^{*i'} ] \\ 
%\label{50}
%$$
Therefore, considering Eqs.~(\ref{43}), (\ref{44}), 
(\ref{47}), (\ref{48}) and (\ref{49}), 
the effective actions of the string amount to 
\beq
S &= & - T\int {d^2 x\sqrt {\det (\delta _{ij}  - \partial _a \phi _i \partial ^a \phi _j )} }  \non
  && + T_{\cal I} \int {d^2 x} \sqrt {|\det g^o |} \frac{1}{{1 + |\Phi |^2 }}
 \left[\partial _a \Phi ^{i'} \partial ^a \Phi ^{*i'}  - \frac{1}{{1 + |\Phi |^2 }}\Phi ^{j'} \Phi ^{*i'} \partial _a \Phi ^{i'} \partial ^a \Phi ^{*j'} \right] \non
 && + k_1 \int {d^2 x} \sqrt {|\det g^o |} K_{ab}^i K^{iab}  \non 
 && + k_2 \int {d^2 x} \sqrt {|\det g^o |} (\partial _a x^\mu  \partial ^a x_\mu  )^2  \non
 && + k_3 \int {d^2 x} \sqrt {|\det g^o |} \partial _a x^\mu  \partial ^a x_\nu  \partial _b x_\mu  \partial ^b x^\nu  \non 
 &&+ k'_1 \int {d^2 x} \sqrt {|\det g^o |} g^{\cal I}_{i'j'^*} \partial _a \Phi ^{i'} \partial ^a \Phi ^{*j'} g^{\cal I}_{k'l'^*} \partial _b \Phi ^{k'} \partial ^b \Phi ^{*l'}  \non 
 &&+ k'_2 \int {d^2 x} \sqrt {|\det g^o |} g^{\cal I}_{i'j'^*} \partial _a \Phi ^{i'} \partial ^b \Phi ^{*j'} g^{\cal I}_{k'l'^*} \partial _b \Phi ^{k'} \partial ^a \Phi ^{*l'}  \non 
 &&+ k'_3 \int {d^2 x} \sqrt {|\det g^o |} g^{\cal I}_{i'j'^*} \partial _a \Phi ^{i'} \partial ^b \Phi ^{*j'} g^{\cal I}_{k'l'^*} \partial _b \Phi ^{*l'} \partial ^a \Phi ^{k'} .
\label{51}
\eeq                   
As a summary, the first term corresponds to the vortex low energy fluctuation 
in the target spacetime; the second term describes the interactions between 
the internal orientational moduli and the spatial moduli, which result 
quartic derivative coupling between these two modes. The third term is 
the extrinsic curvature coupling as the characteristics of the embedding 
of the string in the higher dimensions. The last five terms supplement us 
with other quartic derivative couplings with respect to spacetime and internal 
space modes respectively, and $k_i ,k'_{i'}$ are coupling constants. 
Among the coupling constants, $T,k_i$ $(i=1,2,3)$ exist 
in the Abelian vortex-string of $N=1$ while $T_{\cal I},k'_{i'}$ $(i'=1,2,3)$ are peculiar to 
the non-Abelian vortex-string for $N>1$.
              
Here we make a comment on the microscopic derivation of 
those effective coupling constants, 
$T,T_{\cal I},k_i,k'_{i'}$ in the effective Lagrangian
(\ref{51}). 
In principle we can determine those coupling constants 
from a given microscopic Lagrangian. 
Let us consider a Bogomol'nyi-Prasad-Sommerfield 
(BPS) vortex-string 
in the non-Abelian $U(N)$ gauge theory
coupled to $N$ Higgs scalar fields in the fundamental representation 
with the common $U(1)$ charges.  
First, the tension $T$ can be calculated to be $T= 2 \pi v^2$ with 
$v$ the VEV of the Higgs fields \cite{Hanany:2003hp,Auzzi:2003fs}.
Second, the coupling between space-time and internal modes, $T_{\cal I}$, 
called the K\"ahler class, can be determined to be
$T_{\cal I} = 4\pi/g^2 $ with the $U(N)$ gauge coupling constant $g$ 
\cite{Hanany:2004ea,Shifman:2004dr,Eto:2004rz,Gorsky:2004ad}. 
In the case of a non-Abelian vortex-string in 
dense QCD \cite{Balachandran:2005ev,Nakano:2007dr}, 
the tension $T$ is given by 
$T=2\pi v^2 \log \Lambda$ with an infrared cutoff $\Lambda$.
The coupling $T_{\cal I}$ has been calculated recently 
in \cite{Eto:2009bh} as 
$T_{\cal I}=4\pi c/g^2$ with some numerical constant $c$ which 
is not in general unity but depends on the parameters 
in the original Lagrangian. 

On the other hand, higher order terms are 
in general difficult to be determined. 
There have been many attempts 
to calculate $k_i$ $(i=1,2,3)$ 
for the Abelian ($N=1$) local vortex \cite{higher},  
but it seems that many discussions on rigidity or stiffness 
have been given without agreement with each other. 
For the non-Abelian $U(N)$ case, 
there are three more parameters 
$k'_{i'}$ $(i'=1,2,3)$ at this order, which have not been calculated yet.
Our work will provide a general basis for 
a microscopic derivation of the higher order effective action.

Recently non-Abelian vortices have been extended from 
the $U(N)$ gauge group to gauge groups $U(1)\times G$
where $G$ is $SO(N)$, $USp(N)$ \cite{Ferretti:2007rp}
or further arbitrary Lie groups \cite{Eto:2008yi}.
Accordingly, the orientational zero modes become 
$SO(2N)/U(N)$, $USp(2N)/U(N)$ and so on.
An extension to those cases is straightforward since 
the construction of those coset spaces is known \cite{Itoh:1985ha}.

%\vspace{15pt}
%\begin{flushleft}
%{\Large III. General p-Brane Dynamics 
%}
%\end{flushleft}
%\vspace{15pt}
%%%%%%%%%%%%%%%%%%%%%%%%%%%%%%%%%%%%%%%%%%%%%%%%%%%%%%
\section{General $p$-Brane Dynamics \label{sec:3}}

In the previous section, the effective actions 
of the vortex-string ($p=1$ brane) 
that describe both the spacetime and internal space spontaneous symmetry 
breakings have been discussed. To generalize, we consider a general $p$-brane 
topological defect whose formation not only breaks the embedded target spacetime 
symmetry but also spontaneously breaks  the internal target group $G_{\cal I}$ down to an 
invariant subgroup $H_{\cal I}$. Therefore, the dynamics of the $p$-brane would be described 
by the associated Nambu-Goldstone modes corresponding to the collective degrees 
of freedom of the enlarged coset space. 
For instance, domain walls ($p=2$ brane in $D=3+1$) 
have $U(N)$ orientational moduli  
in a %supersymmetric 
$U(N)$ gauge theory with 
$2N$ Higgs scalar fields with properly degenerated masses 
 \cite{Shifman:2003uh}.\footnote{
When the Higgs masses are non-degenerated, 
each domain wall has only $U(1)$ internal modulus 
in addition to translational zero mode \cite{Isozumi:2004jc}.}
Non-Abelian monopoles, 
Skyrmions ($p=0$ branes in $D=3+1$) 
and 
Yang-Mills instantons ($p=-1$ brane in $D=3+1$) 
also have orientational moduli in general.\label{inv}

Consider such a $p$-brane embedded in the 
target $D$ dimensional flat spacetime. As a result, its moduli space 
is given by
\beq
ISO(1,D - 1)/[SO(1,p) \times SO(D - p - 1) ]
 \otimes G_{\cal I} /H_{\cal I}  
\label{52}
\eeq
where the spacetime stability group of the $p$-brane takes the form $SO(1,p) \times SO(D - p - 1)$, namely, the 
brane has Lorentz invariant $SO(1,p)$ symmetry in $p+1$ dimensional spacetime along with the 
rotation invariance in the $D - p - 1$ codimensions, and its long wave fluctuations are 
described by the collective coordinates parameterized by the coset space $\Omega _o$ (see Eq.~(\ref{53})).  
On the other hand, the internal moduli space is given by $\Omega _{\cal I}  = G_{\cal I} /H_{\cal I}$, illustrating that the orientation 
of the brane in the internal space breaks the symmetry group $G_{\cal I}$ down to $H_{\cal I}$, where $G_{\cal I}$ stands for a 
general compact, connected, semi-simple Lie group $\{$formed by unbroken generators $T^J$ and broken 
generators $S^I$$\}$, and $H_{\cal I}$ is the unbroken subgroup $\{$formed by generators $T^J$$\}$. As a result, the coset 
representative elements are parameterized as
\beq
\Omega  = \Omega _o \Omega _{\cal I}  = e^{ix^\mu  P_\mu  } e^{i(\phi ^m Z_m  + u_\mu ^m K_m^\mu  )} e^{i\Phi ^I S^I }  
\label{53}
\eeq
where $x^\mu$ are the coordinates that parameterize the $p$-brane world volume in the static gauge, 
with $\mu  = 0,1,\cdots,p$; $Z_m$ are the broken generators associated with the translational directions transverse to the 
brane, with $m = 1,.\cdots,D - p - 1$; $K_m^\mu$ are the broken generators related to the rotations that mix the brane world volume 
and the codimensional directions. In the sector of the internal space, the generators $T^J$ and $S^I$ have 
the following commutation relations:
\beq
[T,T] \propto T,  {\hspace{3pt}}   
[T,S] \propto S 
\label{54}
\eeq
i.e., the generators $S^I$ form a representation for the subgroup $H_{\cal I}$. Accordingly, the complete 
Maurer-Cartan one-forms can be written as                                 
\beq
\Omega ^{ - 1} d\Omega  &= & (\Omega _o \Omega _{\cal I} )^{ - 1} d(\Omega _o \Omega _{\cal I} ) \non   
&= & \Omega _o ^{ - 1} d\Omega _o  + \Omega _{\cal I} ^{ - 1} d\Omega _{\cal I}  \non 
&= & i(\omega ^{A'} P_{A'}  + \omega _{z_{mi} } Z_m  + \omega _{A'}^m K_m^{A'} \cdots + A^M S^M  + B^N T^N ) 
\label{55}
\eeq                   
with        
\beq
i\omega ^{A'} P_{A'} &= & idx^\mu P_{A'} (\delta _\mu  ^{\hspace{3pt}A'}  + u_{m\mu } (U^{ - \frac{1}{2}} (\cosh \sqrt {4U}  - 1)U^{ - \frac{1}{2}} ))_{mn} u_n^{A'} ) \non 
  && - id\phi _m (U^{ - \frac{1}{2}} \sinh \sqrt {4U} )_{mn} u_n ^{A'} P_{A'}  = idx^\mu  e_\mu  ^{\hspace{3pt}A'} P_{A'};  \non 
 i\omega _{z_m } Z_m  &= & id\phi _n (\cosh \sqrt {4U} )_{nm} Z_m  - idx^\mu  u_{n\mu } (\cosh \sqrt {4U} U^{ - \frac{1}{2}} \tanh \sqrt {4U} )_{nm} Z_m;  \non 
 i\omega _{A'}^m K_m^{A'}  &= & idu_n^{B'} (\sinh \sqrt M M^{ - 1/2} )_{nmB'A'}^{} K_m^{A'};  \non 
 iA^M S^M  &= & id\Phi ^I E_I^M S^M 
\label{56}
\eeq                   
in which $U_{mn}  = u_m^\mu  u_{n\mu }$, and $M_{mn\mu \nu }  = 4(u_{m\mu }^{} u_{n\nu }  - 2u_{m\mu } u_{n\nu }  + u_{l\mu } u_{l\nu } \delta _{mn} )$. 
After eliminating the non-dynamical fields $u_\mu ^m$ setting $\omega _{z_m }  = 0$ 
by the inverse Higgs mechanism \cite{Ivanov:1975zq},
the $p$-brane world volume 
metric is given by \cite{Clark:2007rn}
\beq
g^o_{\mu \nu }  = e_{o\mu }^{\hspace{3pt}{A'}} e_{o\nu }^{\hspace{3pt}{B'}} \eta _{A'B'} 
  = (\eta _{\mu \nu }  - \partial _\mu  \phi _m \partial _\nu  \phi _m ) 
\label{57}
\eeq
On the other hand, the metric of the internal coset manifold has the form
\beq
g^{\cal I}_{JO}  = E_J^N E_O^M L_{NM}   
\label{58}
\eeq
where $L_{NM}$ is the metric imposed on the covariant coordinate differentials for the invariant interval 
under the $H_{\cal I}$ transformation
\beq
ds_{\cal I} ^2  = A^M A^N L_{MN}  = g^{\cal I}_{JO} d\Phi ^J d\Phi ^O   
\label{59}
\eeq
Therefore, considering Eqs.~(\ref{57}) and (\ref{58}), the metrics in the enlarged moduli space give us the effective 
theory of these moduli (NG) fields on the brane world volume. The effective actions of $p$-brane 
dynamics can be written as\footnote{When the internal target  space is K\"ahler, 
there also exist the terms in footnote \ref{inv} in general 
unless the curvature tensors are written 
by the metric tensor as in ${\bf C}P^{N-1}$. 
This is because 
the complex structure $I^i_j$ of K\"ahler manifolds 
can be used to construct invariant terms in the effective action. 
When the target space has further invariant tensors, 
they can be also used to construct further invariant terms 
in the action.
}                               
\beq
S &= & - T\int {d^{p + 1} x\sqrt {|\det (\eta _{\mu \nu }  - \partial _\mu  \phi _m \partial _\nu  \phi _m )|} }  + T_{\cal I} \int {d^{p + 1} x} \sqrt {|\det g^o |} g^{\cal I}_{JO} \partial _\mu  \Phi ^J \partial ^\mu  \Phi ^O  \non 
 & &+ k_0 \int {d^{p + 1} x} \sqrt {|\det g^o |} K^{m\mu \nu } K_{\mu \nu }^m  
    + k_3 \int {d^{p + 1} x} \sqrt {|\det g^o |} R \non 
 & & + k_1 \int {d^{p + 1} x} \sqrt {|\det g^o |} (\partial _\mu  x^{\mu '} \partial ^\mu  x_{\mu '} )^2  \non
 && + k _2 \int {d^{p + 1} x} \sqrt {|\det g^o |} \partial _\mu  x^{\mu '} \partial ^\mu  x_{\nu '} \partial _\nu  x_{\mu '} \partial ^\nu  x^{\nu '}  \non 
& &+ k'_1 \int {d^{p + 1} x} \sqrt {|\det g^o |} g^{\cal I}_{JO} \partial _\mu  \Phi ^J \partial ^\mu  \Phi ^O g^{\cal I}_{KL} \partial _\nu  \Phi ^K \partial ^\nu  \Phi ^L  \non  
& &+  k'_2 \int {d^{p + 1} x} \sqrt {|\det g^o |} g^{\cal I}_{JO} \partial _\mu \Phi ^J \partial ^\nu  \Phi ^O g^{\cal I}_{KL} \partial _\nu  \Phi ^K \partial ^\mu  \Phi ^L  \non
& &+  k'_3 \int {d^{p + 1} x} \sqrt {|\det g^o |} R_{\cal I} +... 
%\non
%g^{\cal I}_{JO} \partial _\mu \Phi ^J \partial ^\nu  \Phi ^O g^{\cal I}_{KL} \partial _\nu  \Phi ^K \partial ^\mu  \Phi ^L  \non
%& &+  k'_4 \int {d^{p + 1} x} \sqrt {|\det g^o |} g^{\cal I}_{JO} \partial _\mu \Phi ^J \partial ^\nu  \Phi ^O g^{\cal I}_{KL} \partial _\nu  \Phi ^K \partial ^\mu  \Phi ^L  \non
%& &+  k'_5 \int {d^{p + 1} x} \sqrt {|\det g^o |} g^{\cal I}_{JO} \partial _\mu \Phi ^J \partial ^\nu  \Phi ^O g^{\cal I}_{KL} \partial _\nu  \Phi ^K \partial ^\mu  \Phi ^L  
\label{60}
\eeq                   
in which $x^{\mu '}$ are coordinates that parameterize the target spacetime, and $\mu ' = 0,1,2,\cdots,D - 1$. 
The second term gives us the 
coupling modes between the internal and the spacetime symmetry breakings. As a result, the general 
$p$-brane effective actions in terms of both the spacetime and the internal $G_{\cal I} /H_{\cal I}$ space breakings have 
the synthesis form of Nambu-Goto-NLS model actions. The third term is the extrinsic curvature 
part, describing the rigidity or stiffness of the $p$-brane and specifying its motion in the target space, 
where $K_{\mu \nu }^m$ are the extrinsic curvatures, given by $K_{\mu \nu }^m  = e_\mu  ^{\hspace{3pt}B'} e_\nu  ^{\hspace{3pt}A'} K_{B'A'}^m $, 
and $\omega _{A'}^m  = dx^\mu  e_\mu  ^{B'} K_{B'A'}^m$ \cite{Bandos:1995zw}. 
In the brane-world scenario,
the quartic derivative couplings are 
very crucial in brane dynamics, signaling the brane stiffness and embedding in the extra dimensions. 
Their phenomenological consequences have been explored in collider physics and cosmology as well \cite{Clark:2007wj}.  
The other terms are higher order corrections to the brane actions up to the order of quartic derivatives. 
Besides, additional fermionic degrees of freedom can also be located on the brane world volume to 
describe its oscillations into the superspace for 
BPS \cite{Hughes:1986dn,Ivanov:1999fwa} 
and non-BPS branes \cite{Clark:2002bh,Liu:2007mq} 
in supersymmetric theories. 

%     Furthermore, the internal moduli space can also be extended to the complex Grassmann manifold, 
%such as the non-Abelian domain wall topological defects which have been extensively studied in 
%higher dimension theories \cite{Isozumi:2004jc}. 
Recently, in Ref.~\cite{Eto:2004ii}, 
a non-Abelian $p=3$ brane topological defect in $D=5+1$
with orientational moduli configurations has also been discussed. 
In brane world 
scenarios, standard model particles should be realized as low energy fluctuations localized on 
the $p=3$ brane world volume. Phenomenologically, the physical consequences of the couplings in 
Eq.~(\ref{60}) for the $p=3$ brane, which produce interactions containing quartic derivatives in addition 
to the extrinsic curvature terms, are open appealing problems to be explored.

%%%%%%%%%%%%%%%%%%%%%%
\section{Conclusion and Discussion \label{sec:4}}

The non-Abelian vortex-strings were found in 
supersymmetric $U(N)$ gauge theories and high density QCD.
They have the orientation zero modes in 
the internal color and flavor space.
In this paper,
we have constructed the low energy effective action 
for a non-Abelian vortex-string, the presence of which 
breaks the translational and 
the internal $SU(N)$ symmetries.
It has been turned out to be the invariant synthesis of 
the Nambu-Goto-${\bf C}P^{N-1}$ model action 
in the lowest order. 
We then have constructed higher order terms related to 
the string rigidity in both 
the space-time and the internal spaces.
Our work will provide a general basis to go on 
the microscopic derivation of the effective action 
of the non-Abelian vortex-string.
We have also constructed the effective action of 
a $p$-brane which breaks the translational and 
the internal $G_{\cal I}$ symmetries at the same time.
In appendix \ref{sec:SF} we give a relation 
between the Nambu-Goto-${\bf C}P^1$ model and
the Skyrme-Faddeev model.

As demonstrated in this paper 
the nonlinear realization method offers 
a powerful tool to construct the low energy 
effective action of a single vortex. 
On the other hand 
multiple vortices admit more ample moduli spaces \cite{Eto:2005yh} 
but symmetry is not enough to determine the effective Lagrangian. 
However the moduli space can be written as 
a symmetric product $({\bf C} \times {\bf C}P^{N-1})^k/S_k$ 
of each vortex moduli space 
when all vortices are well separated. 
So the nonlinear realization may work to some extent in this case 
because the position and the orientation of each vortex 
are approximate Nambu-Goldstone modes 
though exact Nambu-Goldstone modes 
are only overall translation and orientation.

We have studied only bosonic action in this paper. 
In ${\cal N}=2$ supersymmetric gauge theories, 
$U(N)$ non-Abelian vortices are BPS 
states preserving a half of supersymmetry 
\cite{Hanany:2003hp,Auzzi:2003fs}.
Therefore Nambu-Goldstone fermions associated 
with the partially broken supersymmetry 
are localized in the vortex-string. 
The effective theory is expected to become 
an invariant synthesis of 
the Nambu-Goto action 
with partial supersymmetry breaking terms
\cite{Hughes:1986dn} 
and the ${\cal N}=(2,2)$ supersymmetric 
${\bf C}P^{N-1}$ model. 
Higher order terms should be severely restricted by supersymmetry.
On the other hand, 
non-Abelian vortices can become non-BPS states 
in ${\cal N}=1$ supersymmetric theories
with an F-term potential. 
In that case, the effective action should 
be an invariant synthesis of 
the Nambu-Goto-Volkov-Akulov action \cite{Clark:2002bh} 
and the bosonic ${\bf C}P^{N-1}$ model. 
Extensions of our work to those cases remain 
as a future problem.

In this paper we have focused on the case of local vortices 
with the number of flavor equal to the number of color, 
$N_{\rm F}=N_{\rm C}$.
When the number of flavor is larger than the number of color, 
$N_{\rm F}>N_{\rm C}$, vortices have a size modulus and
are called semi-local vortices \cite{Vachaspati:1991dz}. 
In this case, the normalizability of zero modes does not hold automatically; 
the ${\bf C}P^{N-1}$ modes are (non-)normalizable when size modulus 
is (non-)zero \cite{Shifman:2006kd}. 
When zero modes are non-normalizable they do not appear in the 
effective theory on the vortex but rather should be interpreted 
as bulk modes.
Therefore we have to pay attention not only to a symmetry structure 
but also to the normalizability of zero modes.

%%%%%%%%%%%%%%%%%%%%%%%%%%%%%%%%%%%%%%%%%%%%%%%%%
%\vspace{15pt}
\section*{Acknowledgments}
\vspace{15pt}
One of the authors (L.X.Liu) would like to thank T.K.Kuo and the THEP group in Physics 
Department at Purdue University for support.  
M.N. would like to thank T.~E.~Clark, S.~T.~Love and 
the THEP group in Department of Physics at Purdue University 
for warm hospitality during his stay where this work was initiated.
The authors also would like to thank Martin Kruczenski for useful comments. 
The work of M.N.~is supported in part by Grant-in-Aid for Scientific
Research (No.~20740141) from the Ministry
of Education, Culture, Sports, Science and Technology-Japan.

%%%%%%%%%%%%%%%%%%%%%%
\begin{appendix}
\section{The ${\bf C}P^1$ Model with Four Derivative Terms}
\label{sec:SF}
The Nambu-Goto-${\bf C}P^N$ Lagrangian (\ref{51}) 
reduces in the case of $N=1$ with $p=3$ to
\beq
S &= & - T\int {d^4 x\sqrt {\det (\delta _{ij}  - \partial _a \phi _i \partial ^a \phi _j )} }  \non
  && + T_{\cal I} \int {d^4 x} \sqrt {|\det g^o |} 
{\partial _a \Phi \partial ^a \Phi^* \over(1+|\Phi|^2)^2}
% \left[\partial _a \Phi ^{i'} \partial ^a \Phi ^{*i'}  - \frac{1}{{1 + |\Phi |^2 }}\Phi ^{j'} \Phi ^{*i'} \partial _a \Phi ^{i'} \partial ^a \Phi ^{*j'} \right] 
\non
 && + k_0 \int {d^4 x} \sqrt {|\det g^o |} K_{ab}^i K^{iab}  \non 
 && + k_1 \int {d^4 x} \sqrt {|\det g^o |} (\partial _a x^\mu  \partial ^a x_\mu  )^2  
 + k_2 \int {d^4 x} \sqrt {|\det g^o |} \partial _a x^\mu  \partial ^a x_\nu  \partial _b x_\mu  \partial ^b x^\nu  \non 
 &&+ k'_1 \int {d^4 x} \sqrt {|\det g^o |} 
{\partial_a \Phi \partial^a \Phi^{*} 
\partial_b \Phi \partial^b \Phi^{*}  \over(1+|\Phi|^2)^4}
 \non
 && + k'_3 \int {d^4 x} \sqrt {|\det g^o |} 
{\partial_a \Phi \partial^a \Phi \partial_b \Phi^* \partial^b \Phi^* \over(1+|\Phi|^2)^4}.
\label{eq:CP^1}
\eeq        
In this case we have noticed that the term with $k_2'$ is degenerate with $k_1'$ term.
%with the ${\bf C}P^1$
%\beq
% g_{i'j'^*} = {1 \over(1+|\Phi|^2)^2}
%\eeq

Here we would like to compare this Lagrangian with 
the Skyrme-Faddeev model \cite{Faddeev:1996zj}, 
which allows knot like solitons in 
3+1 dimensions.
Let us introduce the auxiliary gauge field and its field strength as the following
\beq
&& A_{a} 
 = {-i (\Phi^* \partial_{a} \Phi - \partial_{a} \Phi^* \cdot \Phi) 
 \over 2 (1 + |\Phi|^2)} \\
&& F_{ab} = \partial_{a} A_{b} - \partial_{b} A_{a} =
  {i(\partial_{a} \Phi^* \partial_{b} \Phi 
- \partial_{b} \Phi^* \partial_{a} \Phi)
 \over (1+|\Phi|^2)^2 }.
\eeq
Then the so-called the Skyrme-Faddeev term \cite{Faddeev:1996zj} 
can be written as the field strength squared
\beq
 F_{ab}^2 = 2 {(\partial_a \Phi^* \partial^a \Phi)^2 
 - |\partial_a \Phi \partial^a \Phi|^2 
 \over (1+|\Phi|^2)^4} .
\eeq
This term is also called the baby Skyrme term in 
$d=2+1$ \cite{Piette:1994ug}. 
There is the other independent fourth order term 
\cite{Gies:2001hk,Ferreira:2008nn,Freyhult:2003zb}
which is simply the kinetic term squared: 
\beq
 {(\partial_a \Phi \partial^a \Phi^*)^2 
  \over (1+|\Phi|^2)^4}.
\eeq
This term arises when one constructs 
the low energy effective theory of pure $SU(2)$ Yang-Mills theory
\cite{Gies:2001hk}. 
It is also needed for supersymmetric generalization 
of the Skyrme-Faddeev term \cite{Freyhult:2003zb}.
Therefore the fourth order Lagrangian becomes in total as
\beq
{\cal L}_4 =  K_1 F_{ab}^2 + K_2 {(\partial_a \Phi \partial^a \Phi^*)^2 
  \over (1+|\Phi|^2)^4}
=  { (2K_1+K_2)(\partial_a \Phi^* \partial^a \Phi)^2 
  - 2K_1 |\partial_a \Phi \partial^a \Phi|^2 
 \over (1+|\Phi|^2)^4} .
\eeq
Comparing this with Eq.~(\ref{eq:CP^1}) 
we obtain relations $2K_1 + K_2 = k_1'$ and $K_2 = k_3'$.

~From the isomorphism 
${\bf C}P^1 \simeq S^2 \simeq O(3)/O(2)$, 
the ${\bf C}P^1$ model is equivalent to the $O(3)$ model.
In order to see this equivalence, let us introduce a three vector
${\bf n} = (n_1,n_2,n_3)$ by
\beq
 {\bf n} %&=& \Phi^\dagger \vec{\sigma} \Phi
 &=& {1 \over \sqrt{1+|\Phi|^2}} (1,\Phi^*) \vec{\sigma} 
{1 \over \sqrt{1+|\Phi|^2}}
 \left(\begin{array}{c}
 1 \\
 \Phi 
\end{array}\right) \non
 &=& {1 \over 1+|\Phi|^2} (1,\Phi^*)\left( 
\left(\begin{array}{cc}
0 & 1 \\
1 & 0
\end{array}\right),
\left(\begin{array}{cc}
0 & -i \\
i & 0
\end{array}\right),
\left(\begin{array}{cc}
1 & 0 \\
0 & -1
\end{array}\right)
\right) \left(\begin{array}{c}
1 \\
\Phi 
\end{array}\right) \nonumber\\
&=& \left( {\Phi + \Phi^* \over 1+|\Phi|^2}, 
 {-i \Phi + i \Phi^* \over 1+|\Phi|^2}, 
 {1 -|\Phi|^2 \over 1+|\Phi|^2}
   \right) \label{eq:n}
\eeq
which satisfies the constraint 
\beq
 {\bf n}^2 = 1.
\eeq
Conversely, $\Phi$ is the stereographic coordinate, given by 
\beq
 \Phi = {n_1 + i n_2 \over 1 + n_3}
 = {1 - n_3 \over n_1 - i n_2}.
\eeq
The kinetic term becomes 
\beq 
 {\partial_a \Phi \partial^a \Phi^* \over (1+|\Phi|^2)^2}
 = {1\over 2} \partial_a {\bf n} \cdot \partial^a {\bf n} ,
\eeq
and the field strength can be rewritten as
\beq
 F_{ab} 
 = {\bf n} \cdot (\partial_{a} {\bf n} \times \partial_{b}{\bf n}).
\eeq
Therefore the Skyrme-Faddeev term and the other four derivative term become
\beq
&& F_{ab}^2 
%= 2 {(\partial_a \Phi^* \partial^a \Phi)^2 
% - |\partial_a \Phi \partial^a \Phi|^2 
% \over (1+|\phi|^2)^4} 
 = ({\bf n} \cdot \partial_a {\bf n} \times \partial_b {\bf n})^2
 = (\partial_a {\bf n} \times \partial_b {\bf n})^2 , \\
&& {(\partial_a \Phi \partial^a \Phi^*)^2 \over (1+|\Phi|^2)^4}
 = {1\over 4} (\partial_a {\bf n} \cdot \partial^a {\bf n})^2 ,
\eeq
respectively. 
The total four derivative term ${\cal L}_4$ can be rewritten as
\beq
 {\cal L}_4 
 = K_1 (\partial_a {\bf n} \times \partial_b {\bf n})^2 
 + {K_2 \over 4} (\partial_a {\bf n} \cdot \partial^a {\bf n})^2.
\eeq

\end{appendix}

%%%%%%%%%%%%%%%%%%

%%%%%%%%%%%%%%%%%%%%%%%%%%%%%%%%%%

\end{document}